\documentclass[12pt,preprint]{aastex}

\slugcomment{Accepted by ApJ Letters}
\date{\today}
\shorttitle{SWIFTJ1756.9$-$2508}
\shortauthors{Krimm et al.}

\begin{document}
\title{Discovery of the Accretion-Powered Millisecond Pulsar SWIFT~J1756.9$-$2508 with a Low-Mass Companion}
\author{ H. A. Krimm\altaffilmark{1,2}, C. B. Markwardt\altaffilmark{1,3}, C. J. Deloye\altaffilmark{4}, 
P.\ Romano\altaffilmark{5,6}, D. Chakrabarty\altaffilmark{7}, S. Campana\altaffilmark{5}, J. R. Cummings\altaffilmark{1,8}, D. K. Galloway\altaffilmark{9},
N. Gehrels\altaffilmark{10}, J. M. Hartman\altaffilmark{7}, P. Kaaret\altaffilmark{11}, E. H. Morgan\altaffilmark{7}, J. Tueller\altaffilmark{10}
} 
\altaffiltext{1}{CRESST and NASA Goddard Space Flight Center, Greenbelt,
  MD 20771; krimm@milkyway.gsfc.nasa.gov}
\altaffiltext{2}{Universities Space Research Association, 10211
Wincopin Circle, Suite 500, Columbia, MD 21044}
\altaffiltext{3}{Department of Astronomy, University of Maryland, College Park, MD 20742}
\altaffiltext{4}{Department of Physics and Astronomy, Northwestern University, 2131 Tech Drive, Evanston, IL 60208}
\altaffiltext{5}{INAF--Osservatorio Astronomico di Brera, Via E.\ Bianchi 46, I-23807 Merate (LC), Italy}
\altaffiltext{6}{Universit\`a{} degli Studi di Milano, Bicocca, Piazza delle Scienze 3, I-20126, Milano, Italy}
\altaffiltext{7}{Department of Physics and Kavli Institute for Astrophysics and Space Research, Massachusetts Institute of Technology, Cambridge, MA 02139}
\altaffiltext{8}{Joint Center for Astrophysics, University of Maryland, Baltimore County, 1000 Hilltop
Circle, Baltimore, MD 21250}
\altaffiltext{9}{Centenary Fellow, School of Physics, University of Melbourne, Australia}
\altaffiltext{10}{NASA Goddard Space Flight Center, Greenbelt, MD 20771}
\altaffiltext{11}{Department of Physics and Astronomy, University of Iowa, Iowa City, IA  52242}

\begin{abstract}

We report on the discovery by the {\em Swift Gamma-Ray Burst Explorer} of the eighth known transient accretion-powered millisecond pulsar, SWIFT J1756.9$-$2508, as part of routine observations with the {\em Swift} Burst Alert Telescope hard X-ray transient monitor.  The pulsar was subsequently observed by both the X-Ray Telescope on {\em Swift} and the {\em Rossi X-Ray Timing Explorer} Proportional Counter Array. It has a spin frequency of 182 Hz (5.5 ms) and an orbital period of 54.7 minutes.  The minimum companion mass is 
between 0.0067 and 0.0086 $M_{\odot}$, depending on the mass of the neutron star, and the upper limit on the mass is 0.030  $M_{\odot}$ (95\% confidence level).   Such a low mass is inconsistent with brown dwarf models, and comparison with white dwarf models suggests that the companion is a He-dominated donor whose thermal cooling has been at least modestly slowed by irradiation from the accretion flux.  No X-ray bursts, dips, eclipses or quasi-periodic oscillations were detected.  The current outburst lasted $\approx13$\ days and no earlier outbursts were found in archival data. 

\end{abstract}

	\section{Introduction}

It is now well established that a subset of neutron stars in low-mass X-ray binary systems are millisecond pulsars powered by accretion from either a brown dwarf or degenerate companion in a very close binary configuration.  Such systems are believed to be progenitors of millisecond radio pulsars \citep{alpar82}.  Before the discovery of SWIFT J1756.9$-$2508, there were seven known accretion-powered millisecond pulsars \citep[see][for a full listing]{kaar06}.  The sample is now becoming large enough to start to study classes of these objects which can in turn lead to a better understanding of the properties of neutron star X-ray binaries and their relationship to millisecond radio pulsars.   The known accretion-powered millisecond pulsars all appear to be in very close orbits (with orbital periods ranging between 41.1 and 256.5 min).  For all but one such system, 
XTE J1814$-$338 \citep{bhat05}, orbital constraints rule out a hydrogen-burning main sequence companion, and for four systems, even a brown dwarf companion is not consistent with the orbital parameters.  The application of recent models of white dwarf donors in ultracompact binaries \citep[e.g.][]{delo03,delo07}, shows that, within certain temperature ranges, white dwarfs are a viable model for the companion in the most compact of the X-ray millisecond pulsar systems.

In this Letter, we report the discovery of SWIFT J1756.9$-$2508 (hereafter referred to as J1756), with a frequency of 182.07~Hz.  It was discovered by the {\em Swift Gamma-Ray Burst Explorer} Burst Alert Telescope (BAT) as part of the hard X-ray transient monitor program.  In \S~\ref{data} we discuss the observations made with {\em Swift} and the {\em Rossi X-ray Timing Explorer} (RXTE).  In \S~\ref{timing} and \S~\ref{spectra} we discuss the timing and spectral analysis and in \S~\ref{discussion} we derive limits on the companion mass and compare this source to the other known accreting X-ray millisecond pulsars.  All errors are at the 90\%-confidence level unless otherwise stated.
	\section{Observations and data reduction}\label{data}

\subsection{Swift}\label{bat}

SWIFT J1756.9$-$2508 was discovered using the {\em Swift} BAT hard X-ray transient monitor \citep{atel904}.  The BAT is a large field of view instrument that continually monitors the sky to provide the gamma-ray burst trigger for {\em Swift}. On average more than 70\% of the sky is observed on a daily basis. Sky images are processed in near real-time to detect astrophysical sources in the 15-50 keV energy band\footnote[12]{See http://swift.gsfc.nasa.gov/docs/swift/results/transients/ for details on the transient monitor.}.  

\begin{figure} 
\plotone{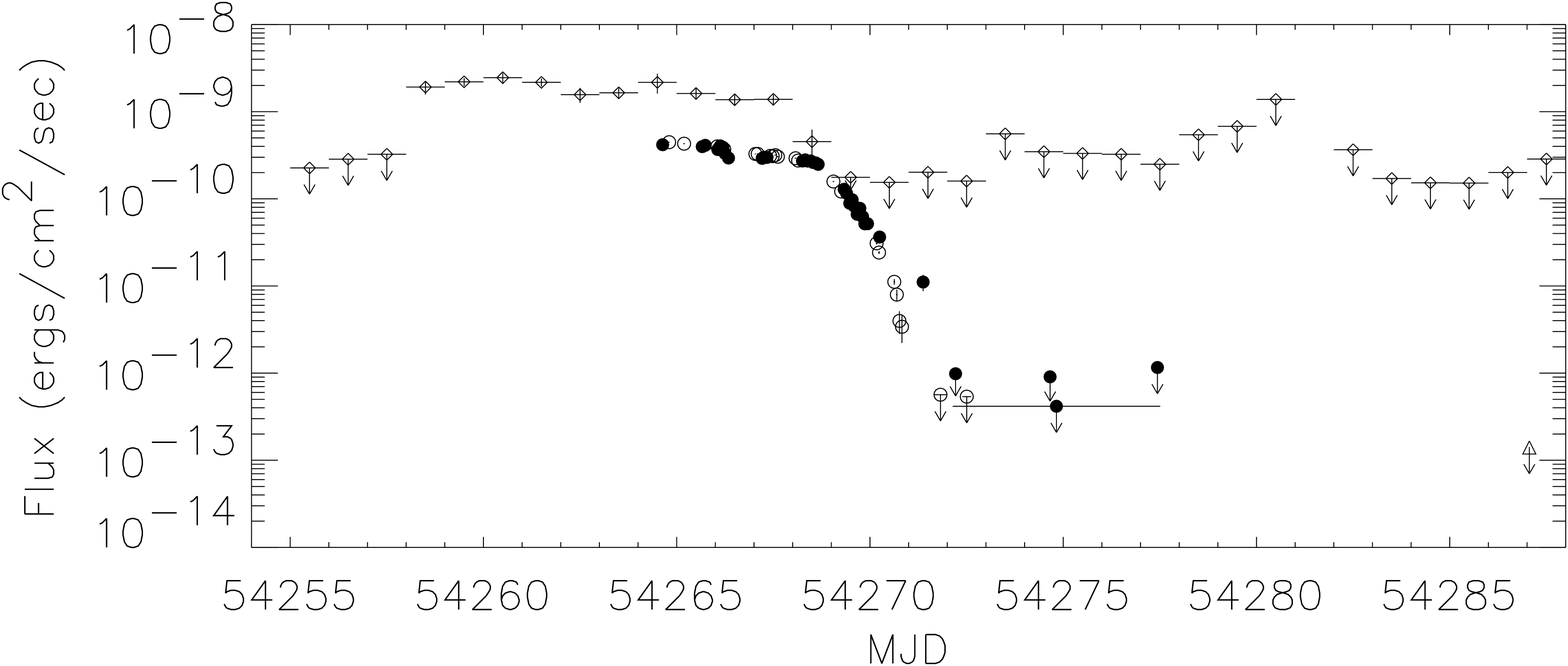}
\caption[XRT-PCA light curve.]{The combined light curve of SWIFT J1756.9$-$2508 from {\em Swift} BAT ({\em diamonds}), {\em Swift} XRT ({\em filled circles}), {\em RXTE} PCA ({\em open circles}), and Chandra \citep[triangle;][]{atel1133}.  The BAT points show flux in the 15--50 keV band, Chandra the 0.3--8.0 keV band, and the other points the 2--10 keV band.  The PCA data is contaminated by emission from the Galactic ridge.  To correct for this, a constant baseline of $1.24 \times 10^{-10}\ {\rm erg\ cm^{-2}\ s^{-1}}$\ was empirically determined and subtracted from the PCA data.   Data points with $< 2\sigma$\ significance are shown as upper limits.  For comparison with the archival upper limits quoted in \S 2.2, for the PCA (BAT) 1 mCrab $\approx 2.4\ (5.2) \times 10^{-11}\ {\rm erg\ cm^{-2}\ s^{-1}}$.
\label{pca_lc}}
\end{figure}

J1756 was detected in the daily full-sky mosaics that are produced in the transient monitor.  It first rose to detectability (Fig.~\ref{pca_lc}) on 2007 June 7 (MJD 54258) with an average rate in {\em Swift} BAT of $0.0083 \pm 0.0015\ {\rm ct\ s^{-1}\ cm^{-2}}$, corresponding to approximately 37 mCrab.  The discovery was announced on June 13 \citep{atel1105}.  
The BAT data for all pointings when the source was fully coded (visible to all BAT detectors) were processed using the {\em Swift} analysis tools, which included cleaning out the diffuse background and flux from other bright sources in the field of view.  

The discovery of J1756 triggered a target of opportunity (ToO) observation in the {\em Swift} X-ray Telescope (XRT) beginning at 15:25:32 UT, 2007 June 13.  It was intermittently observed over 13 days for a total exposure of 60 ks. 
The XRT data were processed with standard procedures ({\tt xrtpipeline} 
v0.10.6 within FTOOLS in the {\tt Heasoft} package v.6.1.2), 
filtering, screening, and grade selection criteria (\citealt{Burrows2005:XRT}).
The photon counting (PC) data collected during the first three segments were 
corrected for pile-up and for the presence of single-reflection rings 
due to the nearby bright source GX~5$-$1.\footnote[13]{XRT observations after MJD 54278 are not usable due to severe contamination from GX~5$-$1.} For the  windowed timing data 
we extracted source events in a square region with a side of 20 pixels. 
Ancillary response files were generated with {\tt xrtmkarf}  
and account for different extraction regions, vignetting and 
 Point-Spread Function 
(PSF)  corrections. 
The light curve (Fig.~\ref{pca_lc}) was corrected for PSF losses and for vignetting.  The best fit XRT 
position is shown in Table~\ref{tab-orbit}.

\subsection{RXTE}\label{rxte}

The discovery of J1756 also triggered a ToO observation in the {\em Rossi X-ray Timing Explorer} (RXTE), starting at 18:42:00 UT, 2007 June 13.   Because GX 5$-$1 is within the
$\approx 1^\circ$ field of view of the {\em RXTE} Proportional Counter Array (PCA) collimator, the PCA pointing
direction was offset by about $18^\prime$, primarily eastward of the
true pulsar position.  The PCA light curve (Fig.~\ref{pca_lc}) has
been corrected to effective on-axis values using the collimator
response function, and spectral response functions account for the
offset as well.  The offset pointing direction is within $3^\prime$ of
the Galactic ridge, and so contamination from diffuse emission is
present.  The amount of contamination was found by examining other
nearby PCA observations, when J1756 was quiescent, and fitting
with a simple phenomenological model (and assuming it is constant).
The contaminating flux was about $\approx$3 mCrab.  The total exposure
was $\approx$104 ks in two observing programs.  

Archived monitor data from {\em RXTE} and {\em Swift} were searched for earlier outbursts of J1756.  The PCA performs regular monitoring observations of the Galactic bulge \citep{swank01} and 
a search of archived PCA bulge scan data shows that the source has not been detected in the 2--10 keV band over the past 8.4 years, with an upper limit of about 5 mCrab (2--10 keV).\footnote[14]{The 2007 outburst of J1756 was not seen in the PCA bulge scan because {\em RXTE} spacecraft slewing constraints prevented observations in early June.  The most recent observation before the outburst was 2007 June 6.}  
This limit is likely to be systematics driven, based on the nearby bright source GX 5$-$1.   
Archived daily BAT monitor data (15--50 keV) were searched back 2.5 years and the {\em RXTE} All-Sky Monitor light curves (2--10 keV) were searched back 11.4 years with no evidence of the source in either instrument with $3\sigma$\ upper limits of $\approx 15$\ mCrab (BAT) and $\approx 50$\ mCrab (ASM).
Due to observing constraints (mostly for sun avoidance), the overall good coverage of the three detectors is $\approx 87$\%.  With these gaps and the fact that even the 2007 outburst was not seen strongly in the ASM, it is possible that an earlier outburst was missed.
Thus we set a tentative recurrence time scale of $\gtrsim 10$\ yr.
 
\subsection{Other Observations}\label{other}

The discovery announcement led to several multiwavelength observations of J1756.  A possible fading near-IR counterpart was announced with $K_s$ = 19.7 on MJD 54270 and $K_s$ = 21.0 on MJD 54282 \citep{atel1132}.  However an observation with the Westerbork Synthesis Radio Telescope during the X-ray outburst  \citep[1.4 GHz; MJD 54266.103;][]{atel1129}  showed no sign of coherent radio pulsations.   Furthermore, J1756 was not detected in observations after the outburst in either 8.7 GHz radio \citep[MJD 54277.371,][]{atel1128} or 0.3--8.0 keV X ray \citep[Chandra upper limit of $1.4 \times 10^{-13}\ {\rm erg\ cm^{-2}\  s^{-1}}$, starting MJD 54287.059,][]{atel1133} observations.
	\section{Timing Analysis}\label{timing}

Fig.~\ref{pca_lc} shows that the outburst of J1756 had a very rapid rise ($\lesssim 1$\ d). 
After this the BAT flux was nearly constant for the next several days.  For the first several days of XRT and PCA observations (up to MJD $\approx 54268.6$), the flux in both instruments showed a slow decline, which can be fit to a nearly exponential decay, $e^{-t/\tau}$, where $\tau = $\ 7.3 $\pm$ 0.4 d.  This matches fairly closely the decay in the BAT light curve.  After this date the flux in all three instruments 
began a steeper decline ($\tau = $\ 0.6 $\pm$ 0.3 d) and was undetectable even in the PCA after MJD 54271, a mere 13 days after the start of the outburst.  This light curve is similar in form to those of three other X-ray millisecond pulsars:   SAX J1808.4$-$3658 \citep{bild01} and IGR J00291+5934 \citep{gall05}, and in particular XTE J1751$-$305 \citep{mark02}, which showed very similar values of $\tau$ \citep{gier05}.    The three other objects have had more than one outburst, on recurrence time scales of $\sim 2 - 3$\ yr, and all outbursts of XTE J1751$-$305 and IGR J00291+5934 and most outbursts of SAX J1808.4$-$3658 have been short ($\lesssim 15$\ days), although notably the 2000 \citep{wijn01} and 2002 \citep{atel505} outbursts of  SAX J1808.4$-$3658 were longer and more erratic.  Given the non-detection of earlier outbursts, J1756 is alone among the fast-decaying millisecond pulsar systems in not showing recurrence.

We searched for high frequency variability by constructing an
FFT-based power spectrum from the PCA data.
The spectrum showed a significant, narrow excess near 182 Hz  \citep{atel1108},
which was also found in
following observations.  The feature was several mHz wide, which
suggested an X-ray pulsar with orbital Doppler broadening.
Examination of power spectra on short ($\approx$300 s) intervals
confirmed a sinusoidal orbital modulation,
making SWIFT J1756.9$-$2508 the
eighth known accreting millisecond X-ray pulsar, and the one with the longest rotation period.

We performed precision timing analysis on the PCA data using the techniques described
in \citet{mark02}.  X-ray event data were recorded in the
\verb|E_125us_64M_0_1s| mode, and event times were corrected to the
solar system barycenter using the position determined from XRT imaging  (see Table~\ref{tab-orbit}).
The $Z^2$ statistic (or Rayleigh statistic) was computed, and the
orbital/timing parameters were adjusted to produce the largest $Z^2$
power of the pulse fundamental.  This procedure is equivalent to radio
or X-ray pulse time-of-arrival fitting \citep{bucc83}, but with no binning.  
After adjustment, we found no long term or
orbital systematic trends.  The binary orbit model was the \verb|ELL1|
from TEMPO software version 11.010,\footnote[15]{http://www.atnf.csiro.au/research/pulsar/tempo/} which is designed for
nearly circular orbits.  The orbital parameters are shown in
Table~\ref{tab-orbit}, and confirm that this is a very compact and
circular orbit.   The pulsation frequencies from MJD 54264.78 to 54267.07, folded on a trial orbital period clearly establish the orbital frequency modulation. 
The pulsed semi-amplitude is a nearly constant
$\approx$6\% throughout the observations. 
No pulsed signal was seen in the 15--50 keV band from 250~s of BAT event starting starting at MJD 54266.065, with a $2\sigma$\ limit of $\lesssim 13$ \%.
With the present data we can place only weak upper limits
on the pulse frequency derivative ($\dot{f}$), since the time baseline
is quite short.

The data from both the PCA and XRT (which cover 100\% of the orbit) were searched for modulation at the orbital period.
The PCA light curve (2--10 keV) shows
 no modulation at the orbital period, after the long term
trend is subtracted. 
The $2\sigma$\ upper limit to orbital modulations is $< 1.2$~\% from the PCA data.  There is also no significant modulation seen in the XRT data.  Furthermore there are no signs of eclipses, dips or bursts.  We searched for rapid variability by constructing power spectra of 64~s segments of event data, with a Nyquist frequency of 2048 Hz,
and averaged contiguous power spectra.  While some variability is present,
especially below $\sim$200 Hz, a detailed timing study is beyond the
scope of this paper.  We do not find any features above 200 Hz which
are obvious kiloHertz quasi-periodic oscillations (QPOs).  For the
summed outburst, the 95\% upper limits on the fractional r.m.s. are
about 6\% (9\%) for 16 Hz (128 Hz) wide QPOs, and a factor of $\sim$2
larger for a single observation.
	\section{Spectral Analysis}\label{spectra}

We fit the Swift/XRT spectrum from each segment in the 0.5--9\,keV energy range
with different models, which included a single absorbed power law \citep[PL;][]{Wilms2000:tbabs}, and an absorbed black body (BB).
A simple comparison of null hypothesis probabilities clearly favours the PL  with
respect to the BB model (20\% vs.\ 4\%, respectively),
with the black body fit producing systematic residuals both at low and high energies.
The derived  absorbing column density is always in excess
of the Galactic value ($N_{\rm H}^{\rm G}=1.47\times 10^{22}$ cm$^{-2}$,
\citealt{DL90}) and there is no indication of a softening of the spectrum with time.
Representative spectral fits are shown in Table~\ref{tab-spect}.
The source was detected in the PCA up to energies $\approx 30$\ keV, and the spectrum is consistent with an absorbed power law with absorption fixed at  the best-fit value of XRT segment 003. The BAT data from 14 to 195~keV were fit by a simple power law model.  The BAT spectrum is softer than the spectra found from fits to the PCA alone or XRT alone, suggestive of a cut-off to the power law.  However joint fits to the BAT, XRT and PCA spectra are not improved with a cut-off power law model.  

Given the uncertainty in the spectrum at high energies, we estimate the integrated outburst fluence to be $\approx (4.5 \pm 0.8) \times 10^{-4}\ {\rm erg\ cm^{-2}}$ (1--10000 keV).  Using the angular  proximity of the source to the Galactic center to estimate the distance, we can, following \citet{bild01}, estimate the time-averaged mass accretion rate, $\dot{M}_{\rm X}\ =\ (9.3 \pm 1.6) \times 10^{-13} M_{\sun}\ {\rm yr}^{-1} d^2_8 m^{-1}_{1.4} T_{10}^{-1}$, where $d_{8}$\ is the distance is units of 8 kpc, $m_{1.4}\ =\ M_{\rm X}/(1.4\ M_{\sun}$), and we assume a neutron star radius of 10 km. The parameter $T_{10}$ is the recurrence time in units 10 yr, and only $T_{10} \gtrsim 1$\ is allowed by the observations. 

\section{Discussion}\label{discussion}

\begin{figure} 
\plotone{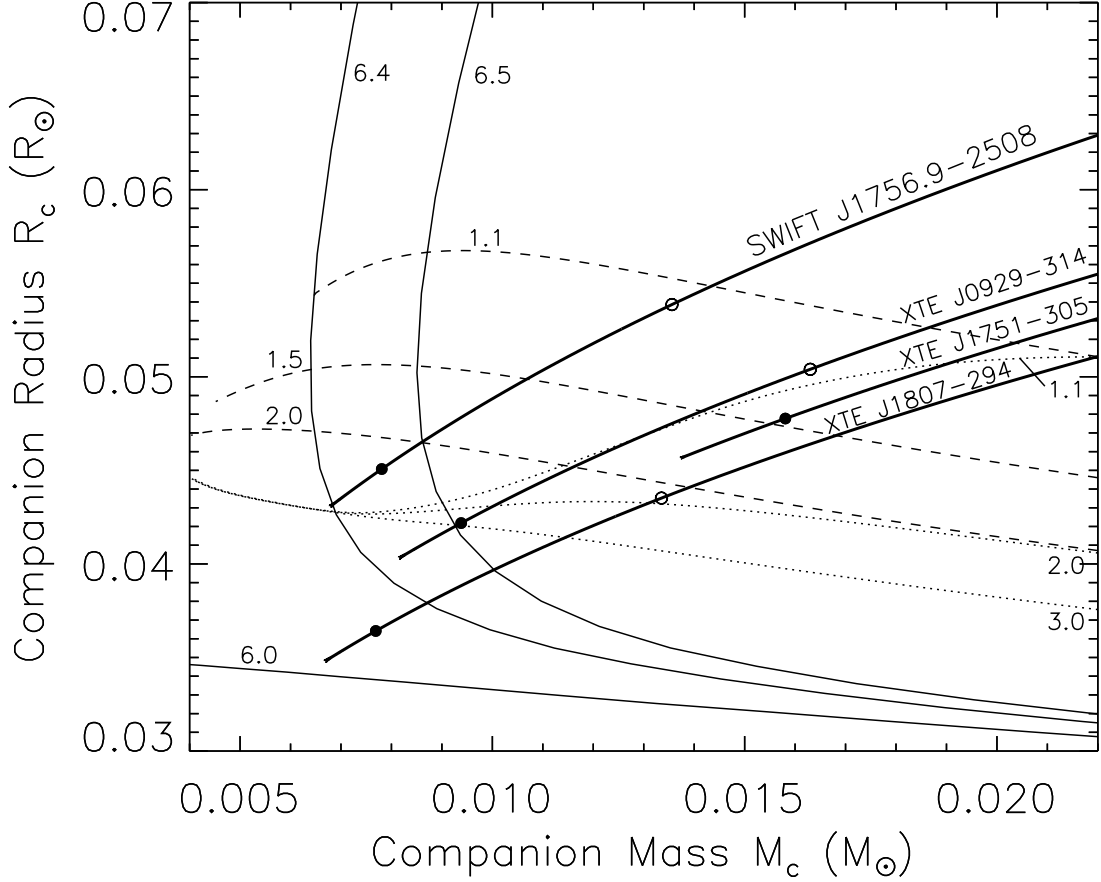}
\caption[Companion mass plot]{Companion radius ($R_C$)  vs. mass ($M_C$) plane, showing the Roche lobe constraints for SWIFT J1756.9$-$2508 and, for comparison, three other X-ray millisecond pulsars with low-mass companions.   The source curves trace possible values of the inclination angle, $i$.  The solid black dots indicate $i = 60^{\circ}$\ and the open black dots indicate $i = 30^{\circ}$.  
The other curves are various white dwarf models.   The dotted and dashed lines show evolutionary sequences with He donor models calculated as in \citet{delo07} (see text for full explanation).
The {\em dotted curves} show cases where no accretion flux is reprocessed.  The {\em dashed curves} show the case where 10\% of accretion flux, as seen by the donor, is thermalized in the donor's atmosphere. 
The numbers next to these curves indicate $\log\psi_{\rm c}$, where higher values of the central degeneracy parameter $\psi_{\rm c}$\ \citep[defined in][]{delo07} indicate a more degenerate and hence compact donor.   
Within each set of models, evolution proceeds to lower $M_C$ and less degenerate donors evolve at larger $R_C$. The {\em solid curves} are for a pure C composition 
from the models of \citet{delo03}.
The number next to each solid curve indicates the log of the temperature in Kelvin used in the model.  
\label{rvm_plot}}
\end{figure}

In combination, the measured mass function and binary orbital period $P_{\mathrm{b}}$ constrain the donor's mass-radius ($M_C$-$R_C$) relation, assuming the donor fills its Roche lobe. For SWIFT J1756.9$-$2508, $P_{\mathrm{b}} = 54.7$\ minutes, a value near the median of the known X-ray millisecond pulsars. 
If we assume a particular value for $M_x$, we can calculate possible values of $M_c$\ from the mass function, $f_x\ =\ (M_c \sin i)^3/(M_x\ +\ M_c)^2$, where $i$\ is the binary inclination to our line of sight and $M_x$\ and $M_c$\ are the masses of the pulsar and companion, respectively.  
The lack of eclipses allows us to set a limit $i < 85^{\circ}$ \citep{pac71}.  For this inclination and  $M_x = 1.4 M_{\odot}$, we have $M_{c,{\rm min}}\ =\ 0.0067 M_{\odot}$.   Assuming a uniform distribution of possible values of $\cos i$\ we can set a 95\% confidence level (C.L.) upper limit on the mass of  $M_{c}\ <\ 0.022 M_{\odot}$ (for $M_x = 1.4 M_{\odot}$).  The range of possible values of $M_c$\ up to this limit and corresponding companion radius $R_c$\ are shown in Fig.~\ref{rvm_plot}. If we consider a larger neutron star of $M_x = 2.2 M_{\odot}$, the limits (95\% C.L.) are  $M_{c}\ <\ 0.030 M_{\odot}$\ and  $R_{c}\ <\ 0.069 R_{\odot}$.  This radius is less than the minimum radius for low-mass hydrogen rich brown dwarfs of any age \citep[][Figure 2]{chab00,bild01}, as expected for a binary period of $\lesssim 80$~min \citep{pac81,nel86}.  Thus we restrict our attention to white dwarf (WD) donor models.

The dotted and dashed curves in Fig.~\ref{rvm_plot} show evolutionary sequences of ultracompact binaries with He donor models originally developed for a study of the AM CVn (double WD) class of ultracompact binaries \citep{delo07}.  
Along these curves, regions of negative slope correspond to adiabatic donor evolution. 
In regions of positive slope the donor is cooling and contracting towards the fully-degenerate He WD $M_C$-$R_C$ relation (see Deloye et al., 2007), essentially traced by the lowest dotted line in Fig.~\ref{rvm_plot}. The donor's cooling is affected by the (very uncertain) efficiency with which irradiating flux generated by the accretion---which always dominates the donor's own flux in these ultracompact binaries---is thermalized in the donor's upper atmosphere. 
The impact of irradiation is to extend the donor's adiabatic evolution phase and to slow its rate of cooling afterwards.

	Since calculations similar to the \citet{delo07} He-donor study do not yet exist for C/O donors, we compare the J1756 constraints to a sample of pure C models of \citet[][{\em solid curves} in Fig.~\ref{rvm_plot}]{delo03}; realistic C/O WDs will lie at somewhat smaller $R_C$. 
These models assume fully-convective interior profiles and do not explicitly treat thermal transport.  However, the physics that produces the transition from adiabatic evolution to cooling in He donors does not depend on composition and C/O donors will also experience a cooling phase.  This means a substantial fraction of the phase space for ``hot'' C/O donors shown in Fig~\ref{rvm_plot} will likely not be evolutionarily accessible (e.g., the nearly vertical regions of the $\log(T_c/\mathrm{K}) >6.4$ ``isotherms''), making a C/O donor in J1756 rather unlikely.  Thus, J1756 likely harbors a He-dominated donor whose thermal cooling has been at least modestly slowed. Irradiation provides a natural mechanism that is able to affect the necessary degree of slowing, as discussed above.
This interpretation places  SWIFT J1756.9$-$2508 in the company of at least two other X-ray millisecond pulsars with very low mass companions: XTE J0929$-$314, \citep{gall02} and XTE J1807$-$294 \citep{fala05}; with its longer orbital period, J1756 provides a unique probe amongst the known ultracompact LMXBs and appears to indicate that irradiation plays a least a modest role in the evolution of the donors in these systems.
 
\acknowledgments 
H. A. K., C. B. M. and J. C. are supported
by the {\em Swift} project. This work is supported at OAB by ASI grant I/011/07/0 and by the Ministry
of University and Research of Italy (PRIN 2005025417).   C. J. D. is supported by NASA through the Chandra X-ray Center, grant number TM7-800. We acknowledge Jean Swank and the {\em RXTE} Science Operations Facility and the {\em Swift} mission operations team for their help in rapidly scheduling these target-of-opportunity observations.


\begin{deluxetable}{lc}
\tablewidth{0pt} 	      	
\tabletypesize{\scriptsize} 
\tablecaption{Timing Parameters of SWIFT J1756.9$-$2508\label{tab-orbit}}
\tablehead{\colhead{Parameter} & \colhead{Value}}
\startdata
Right ascension, $\alpha$ (J2000.0)\tablenotemark{a}  & $17^{\rm h} 56^{\rm m} 57\fs35$ \\
Declination, $\delta$ (J2000.0)\tablenotemark{a}  & $25^{\circ} 06^{\prime} 27\farcs8$ \\
Barycentric pulse frequency, $f_o$ (Hz)                & 182.065804253(72)\tablenotemark{b} \\
Pulsar frequency derivative, $|\dot f|$ (Hz s$^{-1}$)  & $<1\times 10^{-12}$\tablenotemark{c} \\
Projected semimajor axis, $a_x \sin i$ (lt-ms)         & 5.942(27) \\
Binary orbital period, $P_b$ (s)                       & 3282.104(83) \\
Time of ascending node, $T_{\rm asc}$\tablenotemark{d}                  & 54265.28707(6)  \\
Orbital eccentricity, $e$ \tablenotemark{e}                              & $<0.026$ \tablenotemark{c}\\
Pulsar mass function, $f_x$ ($10^{-7} M_\sun$) \tablenotemark{e}         & 1.56(3) \\
Minimum companion mass, $M_c$ ($10^{-3} M_\sun$) \tablenotemark{e}       & 6.7-9.2  \tablenotemark{f} \\
Maximum Power, $Z_{\rm max}^2$                         & 2061 \\
\enddata
\tablenotetext{a}{The estimated position uncertainty is $3\farcs5$. Position is from the XRT and held fixed as a timing parameter.}
\tablenotetext{b}{Uncertainties are $1\sigma$ in the last quoted digits.}
\tablenotetext{c}{95\% upper limit.}
\tablenotetext{d}{Modified Julian days, referred to barycentric dynamical time.}
\tablenotetext{e}{Derived parameter.}
\tablenotetext{f}{For neutron star masses of $1.4 - 2.2 M_\sun$.}
\end{deluxetable}

\begin{deluxetable}{llcccl}
\tablewidth{0pt} 	      	
\tabletypesize{\scriptsize} 
\tablecaption{Sample spectral fits for SWIFT J1756.9$-$2508\label{tab-spect}}
\tablehead{\colhead{ } & \colhead{MJD} & \colhead{Model\tablenotemark{a}} & \colhead{$N_{\rm H}$} & \colhead{$\Gamma$ [$kT$]\tablenotemark{b}} & \colhead{$\chi^2_{\rm red}$/ dof}}
\startdata
XRT & 54266.06 & PL & $5.44 \pm 0.41$ & $2.08 \pm 0.14$ & 1.081/200 \\
XRT & 54266.06 & BB & $2.82 \pm 0.23$ &  [$1.30 \pm 0.05$] & 1.185/200 \\
PCA & 54264.79 & PL & 5.44\tablenotemark{c} & $2.00 \pm 0.14$ & 0.878/47  \\
PCA & 54267.46 & PL & 5.44\tablenotemark{c} & $1.95 \pm 0.08$ & 0.514/48  \\
BAT & 54258 & PL & -- & $2.3 \pm 0.7$ & 0.20/6\\
BAT & 54266 & PL & -- & $2.8 \pm 0.5$ & 0.38/6 \\
\enddata
\tablenotetext{a}{PL=power-law; BB=absorbed black body}
\tablenotetext{b}{Numbers in column 5 are $\Gamma$\ for PL fits and $kT$\ (in square brackets) for the BB fit.}
\tablenotetext{c}{Parameter held fixed.}
\end{deluxetable}

\clearpage

\end{document}